\documentclass[9pt,twocolumn]{article}
\usepackage{graphicx}
\makeatletter
\usepackage{capt-of}
\usepackage{marvosym}
\usepackage{amsmath}
\usepackage{caption}
\usepackage{subcaption}
\usepackage{geometry}
\usepackage{amssymb}
\usepackage{threeparttable}
\usepackage[colorlinks=true, linkcolor=blue, citecolor=blue, urlcolor=blue]{hyperref}
\usepackage[capitalize]{cleveref}
\usepackage{float}
\usepackage{setspace}
\usepackage{multirow}
\usepackage{scalerel}
\newcommand{\CircC}{
\raisebox{-0.21ex}{\scalerel*{\bigcirc}{X}}\kern-0.7em \raisebox{-0.05ex}{\scriptsize C}}
\usepackage[table]{xcolor}
\usepackage{array}
\usepackage{booktabs}
\usepackage{makecell} 
\usepackage{pifont}
\usepackage{rotating}
\usepackage{placeins}
\usepackage{fancyhdr}
\usepackage{orcidlink}
\usepackage[switch]{lineno}

\makeatother

\usepackage[percent]{overpic}
\usepackage[flushmargin]{footmisc}
\makeatletter
\renewcommand{\footnoterule}{
  \kern -3pt
  \hrule width \textwidth height 0.4pt
  \kern 6pt
}
\makeatother

\begin{document}

\twocolumn[
\begin{center}
{\fontsize{15}{17}\selectfont\bfseries Population-Scale Advancing Interface Modeling Reveals How Bacterial Swarms Encode Future Spatial Architecture\par}
\vspace{1.0em}

{\normalsize
Shengyou Duan$^{1}$, 
Zhaoyang Wang$^{2}$, 
Kaiyi Xiong$^{1}$, 
Jin Zhu$^{3,4}$, 
Pengxi Gu$^{1}$, 
Weijie Chen$^{5}$, 
Hongyi Xin$^{6}$, 
Zijie Qu$^{1}$\Letter
\par}

\vspace{0.6em}

{\small
$^{1}$Global College, Shanghai Jiao Tong University, Shanghai, China\\
$^{2}$School of Computer Science and Technology, Beijing Jiaotong University, Beijing, China\\
$^{3}$School of Physics, Georgia Institute of Technology, Atlanta, GA, USA\\
$^{4}$Interdisciplinary Program in Quantitative Biosciences, Georgia Institute of Technology, Atlanta, GA, USA\\
$^{5}$Intelligent Medicine Institute, Shanghai Medical College, Fudan University, Shanghai, China\\
$^{6}$Global Institute of Future Technology, Shanghai Jiao Tong University, Shanghai, China
\par}

\vspace{0.4em}

{\small
\Letter\ Corresponding author. Email:
\href{mailto:zijie.qu@sjtu.edu.cn}{zijie.qu@sjtu.edu.cn}.
\par}
\end{center}

\begin{abstract} Motile bacteria shape microbial function by occupying space, yet how collective motion becomes population-scale architecture remains poorly resolved. Bacterial swarming is not merely surface motion, but a process by which motile populations commit to future macroscopic form. Here, in \textit{Enterobacter} sp.\ SM3, a gut-associated swarmer linked to mucosal repair, we treat the advancing colony--environment interface as a morphodynamic state through which local motility becomes spatial order. We built SwarmEvo across thermal, hydration, and substrate-mechanical conditions and developed Morpher to resolve and propagate interface states. Counterintuitively, within the permissive assay range, condition labels only weakly separated future trajectories, whereas colony-specific interface geometry constrained later expansion, indicating that swarm fate is written into the interface rather than prescribed by condition identity. Boundary fidelity was decisive: a 0.67 percentage-point segmentation gap expanded into a 2.4--3.1 IoU-point forecasting loss. Preserving front displacement, protrusion continuity, and branch memory, Morpher predicted late-stage expansion with 95.42\% mIoU, 10.61 px HD$_{95}$, and 3.93 px ASSD. These results identify the advancing interface as a state-bearing layer through which motility and environmental constraint are converted into future spatial form, enabling disease-relevant microbial organization to be read before endpoint architecture emerges. \end{abstract}

\vspace{2.0em}
]

\section{Introduction}

Microbial function is enacted in space. Across host-associated, environmental, and engineered surfaces, bacteria do not simply accumulate as cell numbers; they partition territory, encounter gradients, compete at interfaces, exhaust local resources, and reshape the physical and chemical landscape that later cells inherit~\cite{best2025metabolic, lotstedt2024spatial, lee2022microbiome, ntekas2026spatial}. At this scale, oxygen access, nutrient exposure, colonization, antimicrobial stress, and tissue-associated ecological reorganization are integrated into colony-scale spatial order~\cite{gude2020bacterial, zhang2010collective, richter2024enhanced}. Bacterial swarming is a particularly direct example of this process. During swarming, motile cells move collectively across semi-solid substrates, generating advancing fronts, transient protrusions, periodic advances, and macroscopic architectures that cannot be inferred from the trajectory of any single cell~\cite{yan2019ultimate, kearns2010field, beer2019statistical, bru2023swarming}. The central biological question is therefore not only how bacteria move, but how cell-scale motility is organized into population-scale spatial form.

Studies of bacterial motility have resolved many of the microscopic mechanisms that make swarming possible, including flagellar dynamics, swarmer-cell differentiation, local interactions, fluid flows, and edge-associated motion~\cite{rauprich1996periodic, rather2005swarmer, ingham2008swarming, kaiser2007bacterial}. These mechanisms are essential, but they do not by themselves explain how a growing population commits to a macroscopic architecture. At the colony scale, the relevant questions change: which regions remain reachable, which protrusions continue to advance, which branches lose access to new substrate, and how local boundary events become routes of territorial occupation. Conventional swarming assays usually reduce this process to an endpoint: a binary motility call, an expansion radius, a growth rate, or a late-stage colony morphology~\cite{mytilinaios2012growth}. Such measurements are useful, but they are retrospective. They record the architecture after the population has already constrained the space it can occupy.

The missing variable is the state of the advancing interface. In an expanding swarm, this interface is the population-scale site at which motility, growth, crowding, substrate contact, and collective organization are translated into territorial advance. New substrate is first sampled there. Curvature alters local access to space. Protrusions create directional bias. Branches preserve the history of earlier advances and arrests. A finger-like sector may continue to elongate, split, merge, or stall, while a near-concentric front may remain coherent or gradually lose symmetry. These boundary events do more than decorate the colony edge. They change which sectors can dominate later expansion and which regions of space will remain inaccessible. This is consistent with the broader principle that, in spatially expanding microbial populations, the frontier is a thin but active layer that shapes large-scale structure, whereas the colony interior can be less informative about future expansion~\cite{hallatschek2007genetic}. Forecasting swarming therefore requires this interface to be treated as a state variable through which local motion is integrated into future macroscopic order.

This distinction matters most in systems where collective motility remodels local habitats. Swarming and related forms of bacterial surface motility can reorganize host-associated and mucus-like environments, influence colonization and competition, and contribute to antimicrobial tolerance and virulence-associated behaviors~\cite{zegadlo2023bacterial, pawul2024mucin, jeckel2023simultaneous, lai2009swarming, overhage2008swarming, piskovsky2023bacterial}. In inflammatory bowel disease (IBD), mucosal injury is accompanied by altered microbial spatial organization, changes in mucus-associated substrates, and the formation of new local ecological niches~\cite{swidsinski2005spatial, fang2021slimy}. \textit{Enterobacter} sp.\ SM3 (SM3), a swarming commensal isolated from the murine gut, has been associated with intestinal microbial reorganization and mucosal repair, whereas swarming-deficient mutants lose these beneficial effects~\cite{de2021bacterial, pollackmilgate2024sm3, chen2021confinement, PhysRevE.109.064402}. Mucin and mucus metabolism can further modulate swarming and gastrointestinal colonization by \textit{Enterobacter}~\cite{pawul2024mucin, sinha2025carbapenem}. For SM3, the key phenotype is therefore not simply the presence or absence of swarming. It is how the interface advances, branches, and claims space.

Experimental and computational methods now allow bacterial colony dynamics to be measured with increasing precision, but most of them still stop short of using the interface as a predictive state. Early image-analysis pipelines showed that colony growth and morphology could be detected, counted, and quantified at scale~\cite{brugger2012automated, zhang2022mlcfu, zhang2022review}. Physical studies of active suspensions and deformable interfaces further showed that collective expansion depends on flows, curvature-dependent growth, local motion, and edge dynamics~\cite{jena2025spatio, xu2023geometrical}. More recent biological image-analysis approaches have extended colony readouts from coarse descriptors to richer microbial phenotypes~\cite{zhang2021deep, litjens2017survey, greener2022guide}, including automated colony detection~\cite{ferrari2017bacterial}, motility-state classification~\cite{paquin2022spatio}, coupled detection--growth analysis~\cite{nagy2023bacterial}, and imaging or sensing readouts of microbial morphology~\cite{wang2020early, doshi2023engineered}. Single-image recognition of swarming probability has also shown that motility phenotypes can be assessed rapidly and objectively~\cite{li2025deep}. These advances establish swarming morphology as measurable. They do not yet establish whether interface morphology expressed at one time point is sufficient to constrain the colony's future path.

We address this gap by modeling the advancing interface of SM3 swarming as a population-scale predictive state. We assemble SwarmEvo, a time-resolved dataset of colony-scale swarming morphologies collected across thermal, hydration, and substrate-mechanical conditions. This design separates three questions that are often conflated in swarming analysis: whether the interface can be measured with boundary-level fidelity, whether nominal assay conditions explain the diversity of future trajectories, and whether the geometry already expressed by an individual colony constrains the macroscopic configuration it will later realize. By recovering boundary-resolved interface states and modeling their subsequent evolution, we test whether future expansion is organized primarily by external condition labels or by the boundary history formed during growth.

Nominal environmental conditions alone do not separate future trajectories. Colony-specific front geometry provides the stronger predictive coordinate, showing that later swarm architecture is already constrained by the interface. Boundary-resolved segmentation is therefore not merely a refinement, but a prerequisite for state estimation: errors in diffuse rims and finger-like protrusions are amplified during autoregressive propagation. By preserving front displacement, protrusion continuity, and structural memory, morphology-aware forecasting predicts late-stage expansion without reducing swarming to a smoothed growth rate. These findings reveal the advancing interface as the active layer where bacterial swarms encode future spatial architecture, converting cell-scale motility and environmental constraint into boundary history that specifies later population-scale form.

\section{Results}

\subsection{Population-scale imaging reveals the advancing interface as a predictive state}

\begin{figure*}[p]
\centering
\includegraphics[width=\textwidth]{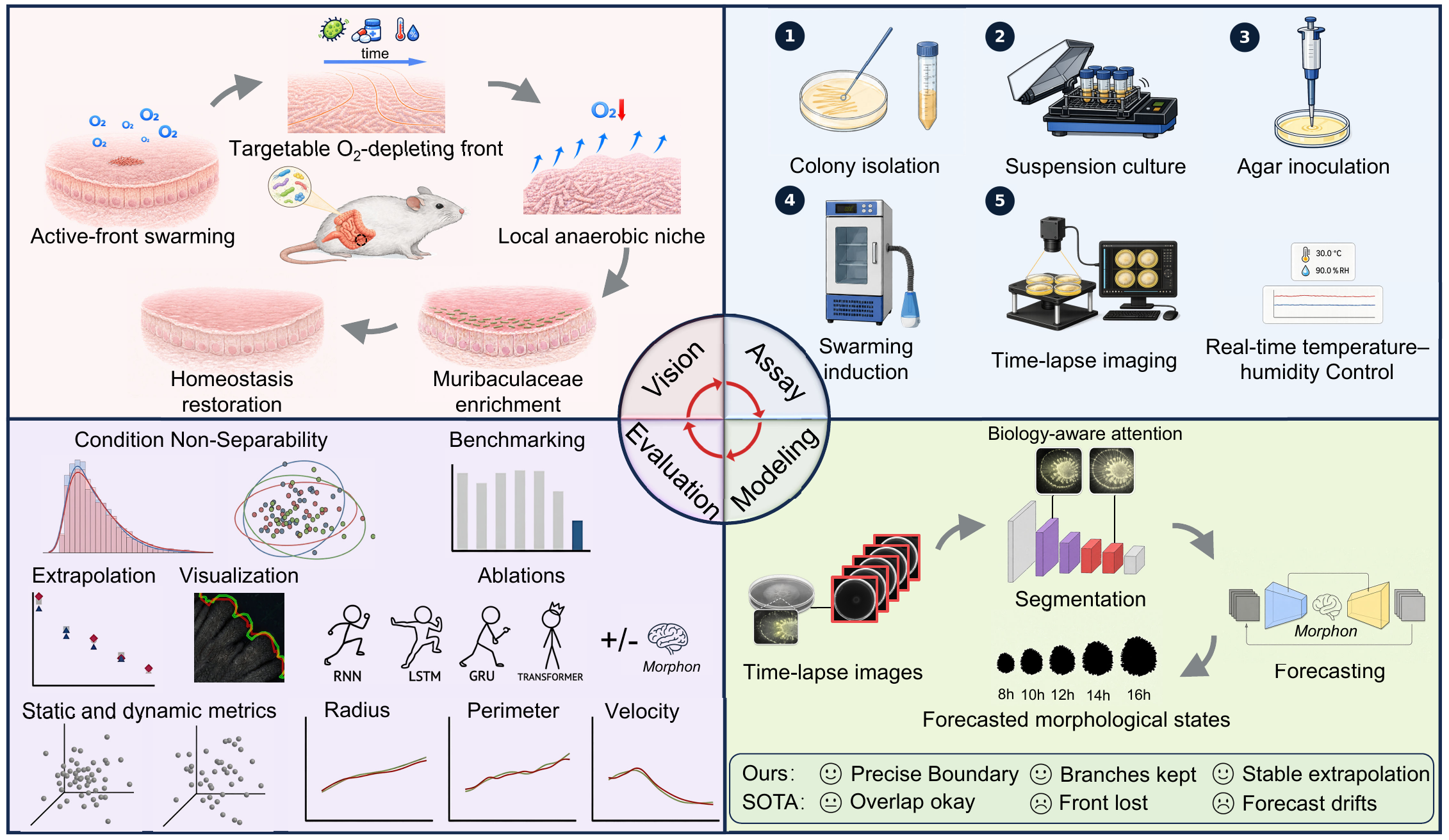}
\caption{
\textbf{Population-scale advancing interface modeling of SM3 swarming expansion.}
The active front of SM3 swarming carries early information about the future macroscopic configuration of the population before that configuration is realized as a final colony architecture. In gut-relevant inflammatory settings, predicting this front could make perturbation spatially targeted rather than endpoint-driven. SM3 colonies were isolated, grown in suspension, inoculated on agar, induced to swarm under controlled temperature--humidity conditions, and recorded by time-lapse imaging. Time-lapse images are converted by Morpher into boundary-resolved colony states that retain diffuse rims and branch-like protrusions, and these states are propagated to forecast future morphology from observed front history. The framework is evaluated through static and dynamic front metrics, forecasting benchmarks, extrapolation tests, visual comparisons, and ablations, linking active-front measurement to geometry-consistent prediction of bacterial swarming fate.
}
\label{fig:f1}
\end{figure*}

Time-lapse imaging showed that SM3 swarming acquires colony-scale order through the advancing interface. Across repeated assays, colonies did not follow a single stereotyped growth path, but occupied a reproducible morphological spectrum with two dominant regimes: an anisotropic branching regime, marked by finger-like advancing fronts, and a near-concentric regime, marked by approximately isotropic radial expansion (Figure~\ref{fig:f1}). The relevant observation was not simply that two morphologies were present. It was that each regime persisted through successive frames as a distinct way of organizing the interface during collective expansion.


This observation changed the object of analysis. Whole-colony images contain illumination, interior texture, agar background, and other assay-dependent appearance cues, whereas the future configuration of the swarm is organized most directly at the interface. New space is first occupied there, and local protrusions, curvature, and branch history alter which parts of the population can continue to expand. We therefore treated the interface, rather than raw image appearance or endpoint morphology, as the state to be measured and propagated. Morpher converted time-lapse images into boundary-resolved colony states that retained diffuse rims and branch-like protrusions, and modeled forecasting as the temporal evolution of these states.

\subsection{Assay conditions do not separate future swarming trajectories}

\begin{figure*}[p]
\centering
\includegraphics[width=\textwidth]{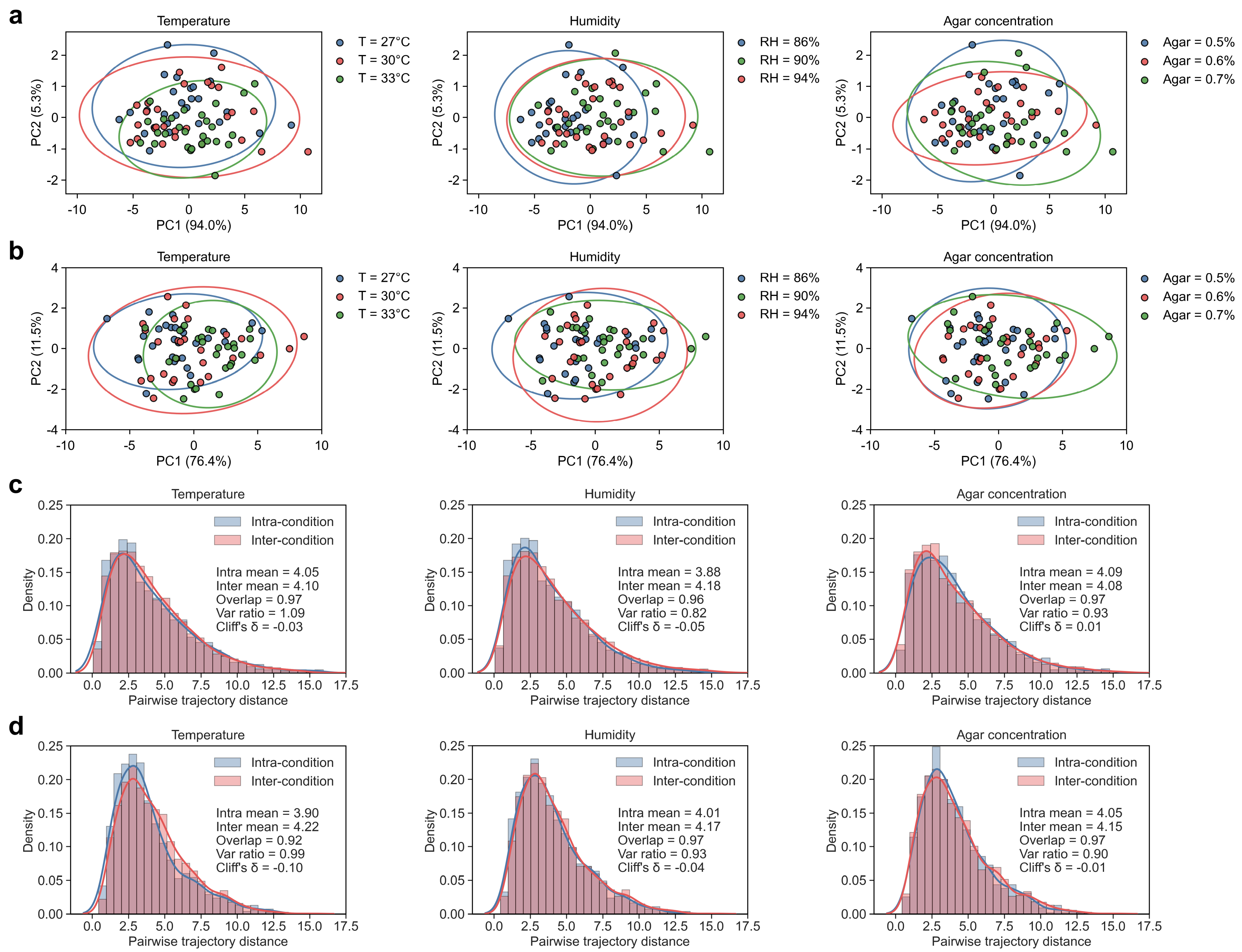}
\caption{
\textbf{Condition non-separability of SM3 swarming trajectories.}
Each colony was represented by a time series of perimeter or area values, and separability was tested under temperature, humidity, and agar-concentration groupings. \textbf{a} PCA of perimeter time-series features. \textbf{b} PCA of area time-series features. \textbf{c} Within- and across-condition pairwise distances between perimeter time series. \textbf{d} Within- and across-condition pairwise distances between area time series. Across both representations, colonies from different assay conditions remain broadly overlapped, indicating that nominal condition labels weakly explain swarming trajectory variation.
}
\label{fig:f2}
\end{figure*}

Nominal assay conditions were insufficient to organize SM3 swarming dynamics. Across temperature, humidity, and agar-defined substrate mechanics, colonies occupied overlapping regions of trajectory state space rather than separating by condition (Figure~\ref{fig:f2}). External assay labels alone therefore provided a weak basis for forecasting future expansion.

This non-separability was evident in principal component analysis of trajectory-level perimeter and area features. Colonies from different temperatures, humidities, and agar concentrations remained broadly intermingled in the leading principal-component spaces (Figure~\ref{fig:f2}\hyperref[fig:f2]{a,b}). Pairwise trajectory distances gave the same result. For both perimeter and area time series, within-condition and across-condition distance distributions nearly coincided, with high overlap coefficients (0.92--0.97) and Cliff's $\delta$ values close to zero across all three condition groupings (Figure~\ref{fig:f2}\hyperref[fig:f2]{c,d}).


Swarming trajectories were therefore not direct readouts of temperature, humidity, or agar concentration. These conditions shaped the arena of expansion, but they did not specify the route by which a colony occupied space. That route was better captured by colony-specific front geometry, motivating prediction from the interface state rather than from condition identity.

\subsection{Boundary-resolved segmentation estimates the interface state}

\begin{figure*}[p]
\centering
\includegraphics[width=0.78\textwidth]{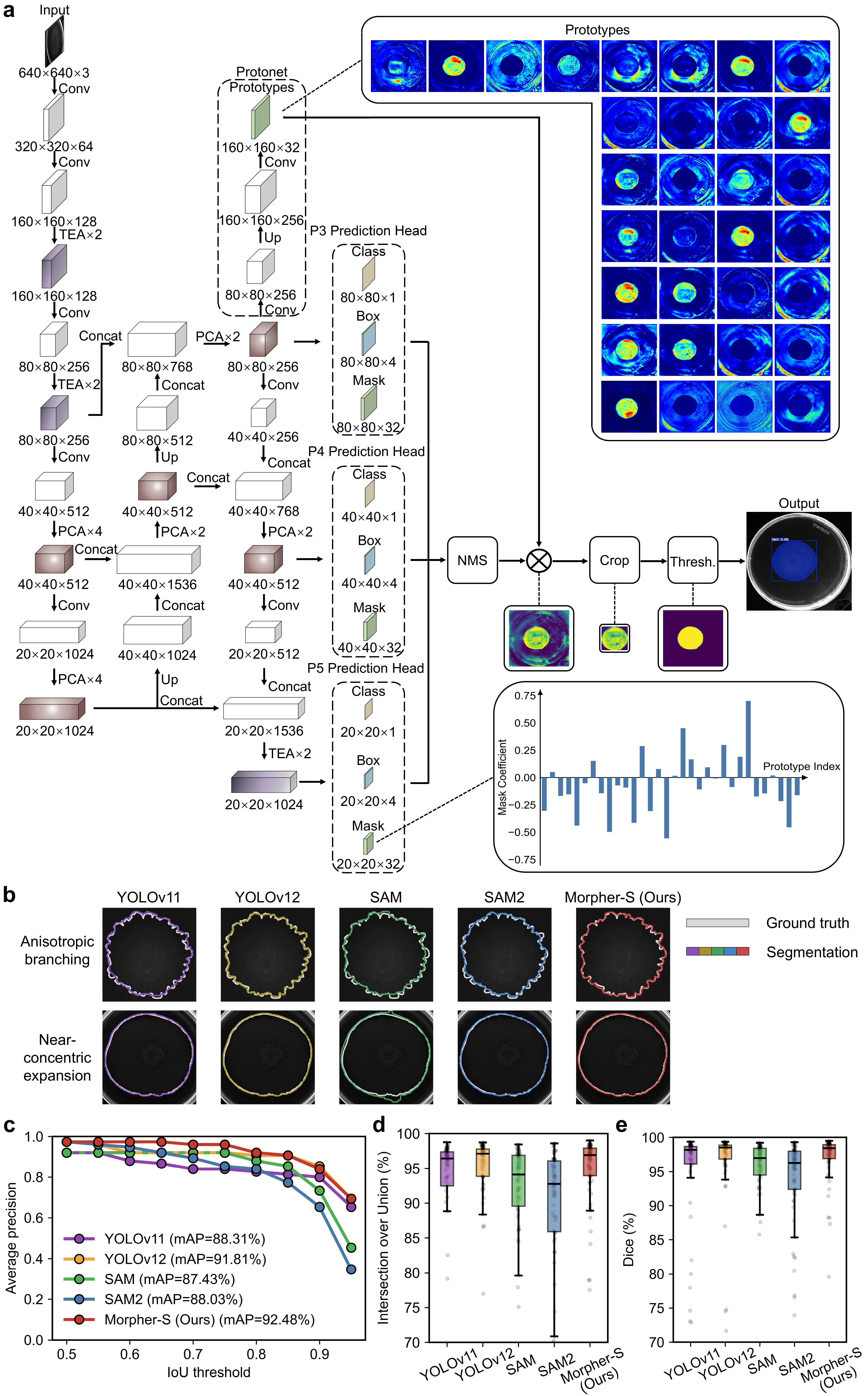}
\caption{
\textbf{Morpher-S estimates the boundary-resolved interface state.}
\textbf{a} Morpher-S is built as a prototype-based instance segmentation model to recover the colony boundary as the state used for forecasting. Texture--Edge Attention preserves weak rim texture and fine edge signals in the backbone, while Polar--Context Attention maintains radial and angular consistency during multi-scale fusion.
\textbf{b} Representative anisotropic branching and near-concentric colonies segmented by competing backbones. Morpher-S preserves distal branches, diffuse rims, and the outer front more faithfully than YOLOv11~\cite{khanam2024yolov11}, SAM~\cite{kirillov2023segmentanything}, and SAM2~\cite{ravi2024sam2}, with YOLOv12~\cite{tian2025yolov12} as the closest competitor.
\textbf{c} AP--IoU curves on the SwarmEvo segmentation test set (78 colonies), showing stronger separation at high IoU thresholds where boundary errors are penalized.
\textbf{d} Image-wise IoU distributions on the same test set.
\textbf{e} Image-wise Dice distributions on the same test set.
}
\label{fig:f3}
\end{figure*}

Once the advancing interface is treated as the predictive state, segmentation becomes a biological measurement rather than a preprocessing step. It determines which boundary history is available for temporal propagation. A smoothed protrusion, a displaced rim, or a contracted branch does not merely reduce pixel-level accuracy; it changes the inferred sequence of territorial advances and arrests. The strongest differences among segmentation backbones appeared at distal protrusions and diffuse rims, especially in the anisotropic branching regime (Figure~\ref{fig:f3}\hyperref[fig:f3]{b}).

This separation was clearest where boundary fidelity was most demanding. YOLOv11 captured the global colony outline but shortened or fragmented slender fingers. SAM and SAM2 preserved the colony core while suppressing distal protrusions, shifting the front toward a smoother and more circular contour. Morpher-S remained aligned with both the outer envelope and fine branches, with YOLOv12 as the closest competitor (Figure~\ref{fig:f3}\hyperref[fig:f3]{b}).


The advantage of Morpher-S increased as evaluation became more boundary-stringent. AP--IoU curves separated rapidly at higher thresholds (Figure~\ref{fig:f3}\hyperref[fig:f3]{c}): Morpher-S achieved mAP$_{50:95}=92.48\%$, followed by YOLOv12 at 91.81\%, whereas SAM and SAM2 dropped to 87.43\% and 88.03\%. The same ordering appeared in image-wise IoU and Dice distributions (Figure~\ref{fig:f3}\hyperref[fig:f3]{d,e}), where Morpher-S concentrated in the high-overlap regime and SAM-based models showed broader low-score tails consistent with missing protrusions and front contraction.

\subsection{Morpher predicts swarming expansion by preserving interface geometry}

\begin{figure*}[p]
\centering
\includegraphics[width=0.80\textwidth]{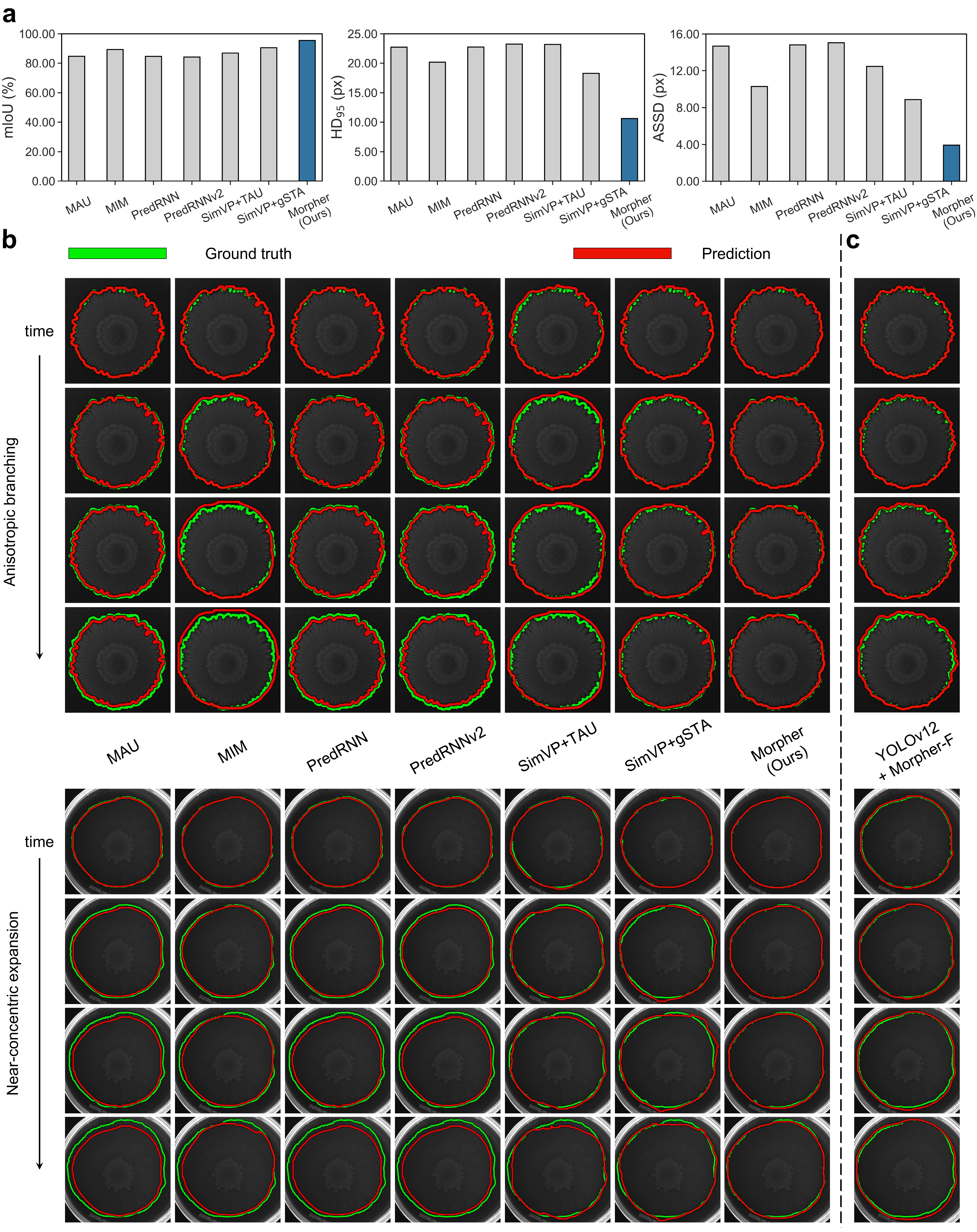}
\caption{
\textbf{Morpher preserves interface state during colony-scale expansion forecasting.}
\textbf{a} Forecasting performance under an 80\% observation / 20\% prediction protocol. Morpher achieves 95.42\% mIoU, 10.61 px HD$_{95}$, and 3.93 px ASSD, giving the best combined overlap and boundary accuracy.
\textbf{b} Long-horizon forecasts on representative anisotropic branching and near-concentric sequences. Generic predictors smooth branches, truncate protrusions, or accumulate boundary drift, whereas Morpher maintains a coherent active front.
\textbf{c} Effect of segmentation state quality on forecasting. Morpher-S masks yield more accurate final-frame predictions than YOLOv12~\cite{tian2025yolov12}, increasing IoU from 94.21\% to 96.63\% in the anisotropic branching regime and from 93.30\% to 96.38\% in the near-concentric regime.
}
\label{fig:f4}
\end{figure*}

Accurate swarming prediction requires the interface to remain correctly positioned, not only regionally overlapped. A forecast can cover nearly the correct area while displacing protrusions, rims, or high-curvature sectors that determine where expansion will continue. We therefore evaluated prediction using HD$_{95}$ and ASSD as boundary-sensitive measures, with mIoU used to quantify coarse occupied area (Figure~\ref{fig:f4}\hyperref[fig:f4]{a}). Boundary displacement is consequential because it moves the predicted interface at which the colony encounters new substrate and reorganizes future space.

We compared Morpher with MAU~\cite{Chang2021MAU}, MIM~\cite{wang2019mim}, PredRNN~\cite{wang2017predrnn}, PredRNNv2~\cite{wang2023predrnn}, SimVP+TAU~\cite{gao2022simvp, tan2023tau}, and SimVP+gSTA~\cite{tan2025simvpv2}, using identical splits and the same 80\% observation / 20\% prediction protocol. Generic predictors often preserved coarse colony extent but remained less accurate at the boundary. MIM and SimVP+gSTA reached mIoU values of 89.32\% and 90.52\%, yet remained substantially worse in HD$_{95}$ and ASSD; PredRNN and PredRNNv2 showed larger boundary errors, while SimVP+TAU remained intermediate.


Morpher performed best across all three metrics, reaching 95.42\% mIoU, 10.61 px HD$_{95}$, and 3.93 px ASSD (Figure~\ref{fig:f4}\hyperref[fig:f4]{a}). Relative to the strongest baseline, SimVP+gSTA, this increased mIoU by 4.90 percentage points and reduced HD$_{95}$ and ASSD by 42.0\% and 55.7\%, respectively. These gains were obtained in the late expansion stage, where the front was extended and small geometric errors could strongly affect the forecast.

Visual forecasts confirmed the metric-level separation (Figure~\ref{fig:f4}\hyperref[fig:f4]{b}). In the anisotropic branching regime, generic predictors smoothed finger-like protrusions, truncated lobe tips, or lagged behind the advancing boundary. In the near-concentric regime, they underestimated radial extent or accumulated boundary drift. Morpher preserved local front perturbations while maintaining global scale, indicating that its advantage lay in preserving the spatial organization of the moving interface rather than only matching occupied area.

\subsection{Interface-state errors amplify during autoregressive propagation}

The interface defines the predictive state, and small errors in its measurement propagate through time. We isolated this effect by holding Morpher-F fixed while varying only the segmentation backbone used to define the input state. The static segmentation difference was modest: Morpher-S achieved 92.48\% mAP$_{50:95}$, whereas YOLOv12 reached 91.81\%. The temporal effect was substantially larger. Over the final 20\% prediction window, YOLOv12-derived masks incurred a 2.4--3.1 IoU-point loss in final-frame prediction.

This amplification appeared in both representative regimes (Figure~\ref{fig:f4}\hyperref[fig:f4]{c}). In the anisotropic branching sequence, final-frame IoU increased from 94.21\% to 96.63\% when Morpher-S defined the state instead of YOLOv12. In the near-concentric sequence, it increased from 93.30\% to 96.38\%. A 0.67 percentage-point difference in static segmentation accuracy was therefore sufficient to induce multi-point degradation in long-horizon prediction.

Small geometric biases were recursively fed back as state during autoregressive boundary evolution, converting local measurement error into cumulative trajectory drift. Thus, boundary-resolved state estimation did not simply affect forecast accuracy. It determined which version of the colony's future expansion the model was allowed to propagate.

\subsection{Ablation reveals distinct requirements for interface propagation and structural memory}

\begin{figure*}[p]
\centering
\includegraphics[width=0.95\textwidth]{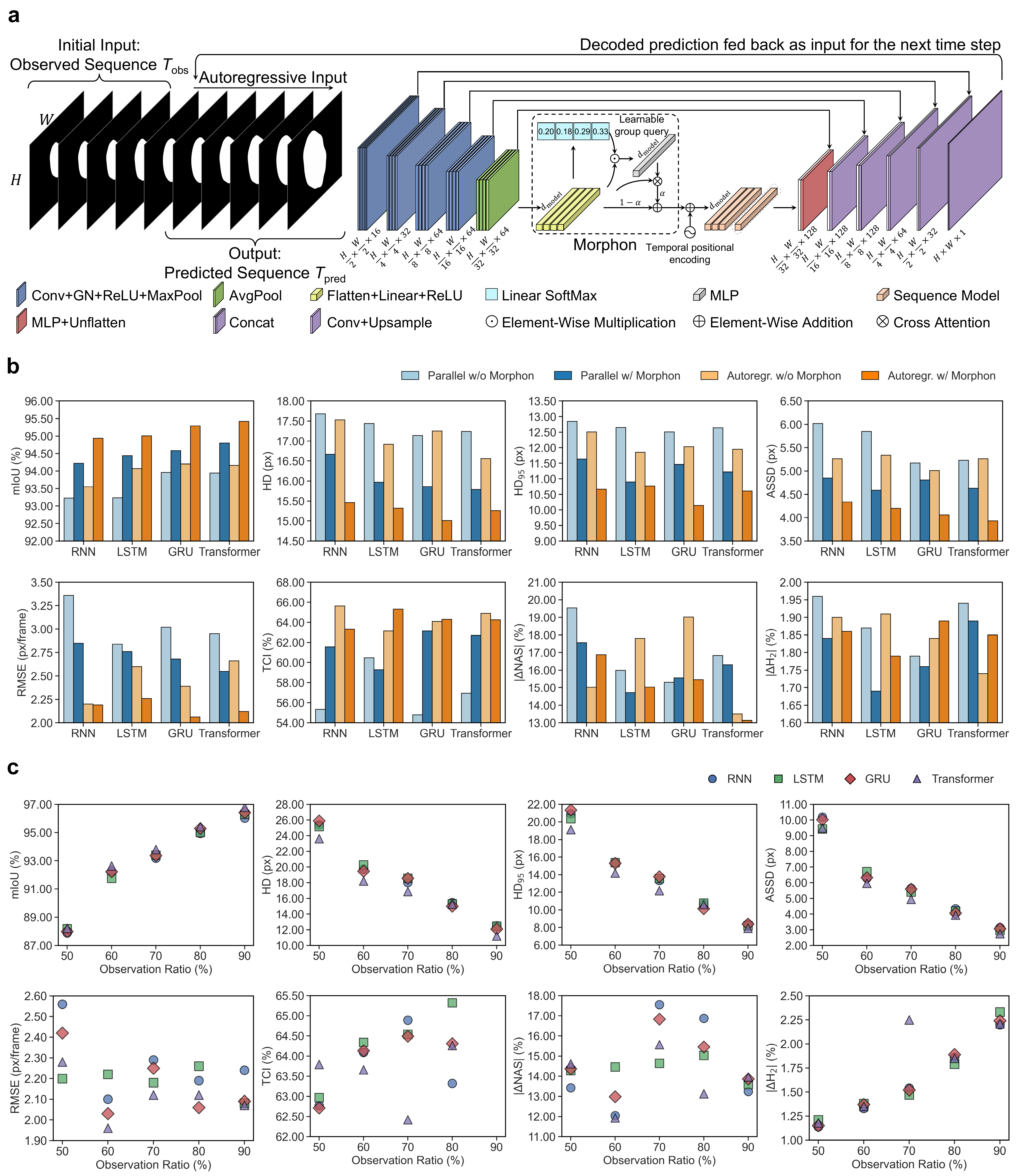}
\caption{
\textbf{Ablation of temporal propagation and structural memory in Morpher-F.}
\textbf{a} Morpher-F architecture. Observed masks are encoded as a morphological latent sequence and propagated autoregressively. Morphon retrieves earlier front states through cross-attention and integrates them through a learnable gate, while the multi-scale decoder reconstructs future boundary geometry.
\textbf{b} Temporal backbones and inference strategies under an 80\% observation / 20\% prediction protocol. Autoregressive decoding improves front propagation, whereas Morphon improves boundary fidelity and anisotropic morphology across RNN, LSTM, GRU, and Transformer backbones.
\textbf{c} Robustness to observation length. Forecasting accuracy and temporal stability increase as the observation ratio rises from 50\% to 90\%, indicating effective use of accumulated front history.
}
\label{fig:f5}
\end{figure*}

Long-horizon forecasting of the interface requires more than placing the boundary near the correct location. The interface carries memory in its branches, curvature, protrusions, and local deformations. A future interface must therefore be correct both in where it is and in how it arrived there. These requirements are related, but not equivalent. We evaluated each model with metrics that separate spatial placement, temporal stability, and interface-structure preservation: mIoU, HD, HD$_{95}$, and ASSD for overlap and boundary accuracy; propagation RMSE and TCI for temporal consistency; and $|\Delta \mathrm{NAS}|$ and $|\Delta \mathrm{H}_2|$ for anisotropic morphology (Figure~\ref{fig:f5}\hyperref[fig:f5]{b}).


Autoregressive decoding improved front propagation across RNN, GRU, LSTM, and Transformer backbones (Figure~\ref{fig:f5}\hyperref[fig:f5]{b}). With the Transformer backbone and Morphon, mIoU increased from 94.80\% under parallel decoding to 95.42\% under autoregressive decoding, while ASSD decreased from 4.63 to 3.93 px. GRU and LSTM followed the same trend. Stepwise state updates preserved local continuity of the moving front, whereas parallel decoding tended to average future shapes into smoother contours.

The ablation separated displacement accuracy from morphology preservation. GRU produced the lowest propagation RMSE, 2.06 px per frame, indicating strong short-term tracking of front motion. It did not, however, produce the best morphology forecast. The Transformer with autoregressive Morphon achieved the strongest overall front reconstruction, with 95.42\% mIoU, 3.93 px ASSD, and the lowest anisotropy deviation, $|\Delta \mathrm{NAS}| = 13.13\%$. Advancing the front and preserving its structure separated across metrics: recurrence reduced short-term displacement error, whereas the strongest morphology scores occurred with autoregressive Morphon.

Morphon improved every backbone, with the largest gains in boundary fidelity and anisotropic structure. In parallel RNN, Morphon increased mIoU from 93.23\% to 94.22\% and reduced ASSD from 6.02 to 4.85 px. In autoregressive GRU, Morphon reduced HD$_{95}$ from 12.03 to 10.14 px. In the Transformer backbone, Morphon reduced ASSD from 5.26 to 3.93 px and lowered $|\Delta \mathrm{NAS}|$ from 16.83\% to 13.13\%. By contrast, $|\Delta \mathrm{H}_2|$ remained below 2.0\% across models, indicating that coarse harmonic shape was less limiting than fine anisotropic organization along the advancing front. These ablations separate the two effects: autoregressive decoding improves front propagation, whereas Morphon improves boundary fidelity and anisotropic-structure preservation.

\subsection{Forecasting stabilizes after colony-specific front geometry emerges}

We varied the observation ratio from 50\% to 90\% using each backbone in its best configuration, with autoregressive decoding and Morphon enabled (Figure~\ref{fig:f5}\hyperref[fig:f5]{c}). This analysis asked when the future macroscopic order of a swarm first becomes readable from its interface. The 50\% setting captured the transition from an early, morphologically similar phase to colony-specific divergence, making it the earliest meaningful point for long-horizon forecasting (Figure~\ref{fig:f6}\hyperref[fig:f6]{d--f}). Before this transition, future expansion was weakly constrained by visible geometry. After it, the interface carried colony-specific information that made prediction substantially more stable.

Boundary and overlap metrics improved steadily with longer observation history. For the Transformer backbone, mIoU increased from 88.22\% at 50\% observation to 96.79\% at 90\%, while ASSD decreased from 9.49 to 2.75 px. The ranking of temporal backbones remained largely unchanged across observation ratios, indicating that architecture, rather than the exact observation--prediction split, dominated performance.

Propagation errors improved most strongly at the onset of colony-specific divergence. RMSE decreased sharply between 50\% and 60\% observation, then remained within a narrow 2.0--2.3 px per-frame range. TCI was similarly stable, clustering around 63--65\% from 50\% to 80\% observation, with the highest value observed for LSTM at 80\% observation (65.32\%). At 90\%, the remaining prediction window was too short for a stable TCI estimate.

Angular metrics revealed the remaining difficulty. Longer observation improved localization, but did not monotonically improve anisotropy preservation. For the Transformer, $|\Delta \mathrm{NAS}|$ decreased from 14.62\% at 50\% observation to 11.93\% at 60\%, rose to 15.57\% at 70\%, decreased to 13.13\% at 80\%, and remained near 14\% at 90\%. $|\Delta \mathrm{H}_{2}|$ showed the same late-stage sensitivity, remaining at 1.18--1.35\% at 50--60\% observation and approaching $\sim$2.2\% at 90\%. Additional history improved front localization, but residual anisotropy error remained tied to the phase alignment of late branch rearrangements.

\subsection{Interface forecasting generalizes across colonies and preserves swarming dynamics}

\begin{figure*}[p]
\centering
\includegraphics[width=\textwidth]{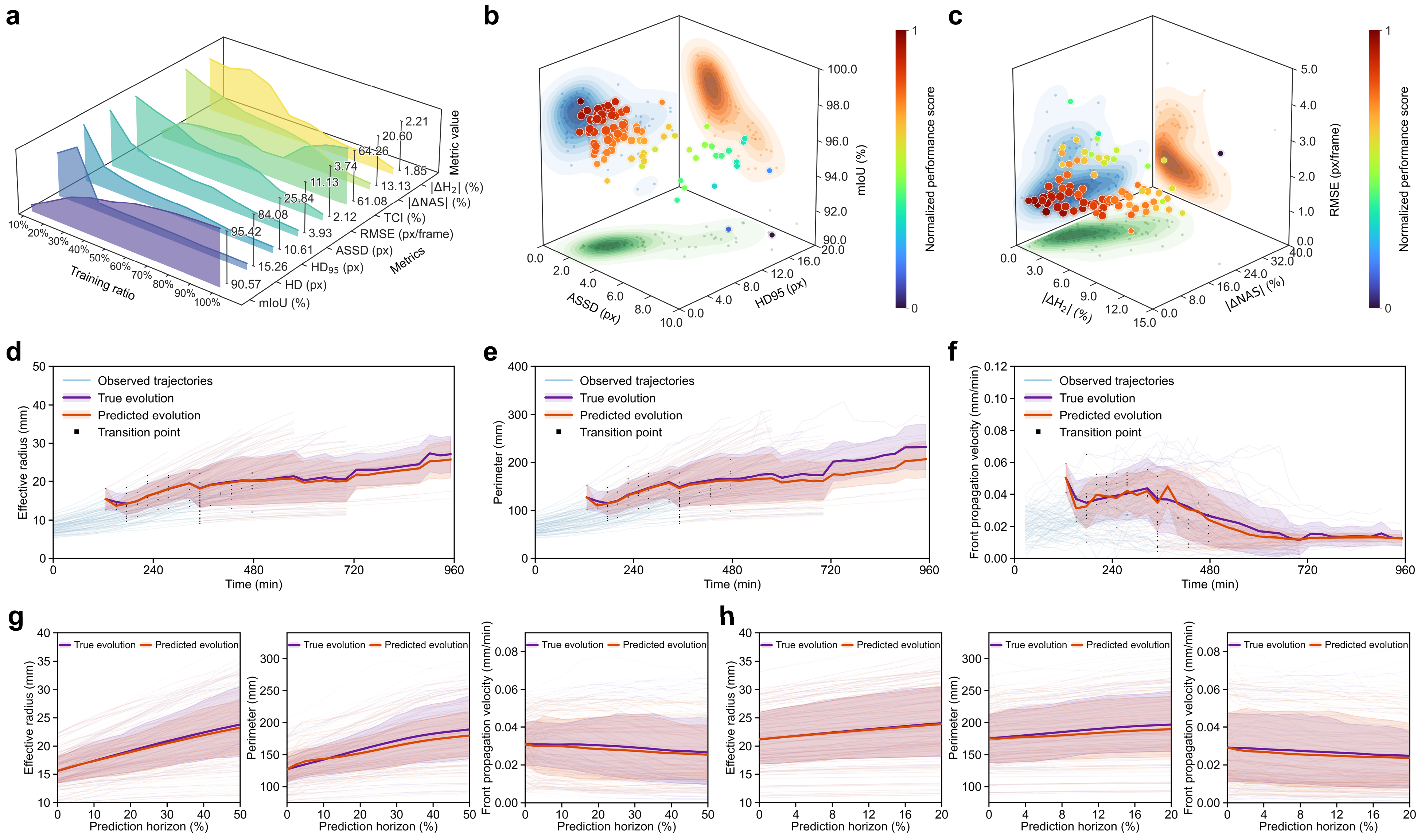}
\caption{
\textbf{Generalization and dynamical consistency of interface forecasting.}
\textbf{a} Effect of training set size on forecasting performance under an 80\% observation / 20\% prediction protocol.
\textbf{b} Leave-one-out forecasting accuracy across 81 colonies, measured by region- and boundary-based metrics.
\textbf{c} Dynamical consistency in leave-one-out evaluation, separating propagation error, anisotropy deviation, and normalized performance.
\textbf{d--f} Time-aligned trajectories of effective radius, perimeter, and front propagation velocity under a 50\% observation / 50\% prediction protocol.
\textbf{g} Prediction-horizon-normalized trajectories under the 50\% / 50\% protocol after alignment of variable-length prediction windows.
\textbf{h} Prediction-horizon-normalized trajectories under the 80\% / 20\% protocol after alignment of variable-length prediction windows.
}
\label{fig:f6}
\end{figure*}

We tested whether interface forecasting remained reliable as the amount of training data was reduced (Figure~\ref{fig:f6}). Training-set-size subsampling showed rapid convergence of performance as more trajectories were included, with only small gains after the dominant modes of swarming variation had been sampled (Figure~\ref{fig:f6}\hyperref[fig:f6]{a}). This pattern argues against simple memorization of individual colonies and supports learning recurrent geometric patterns in interface evolution.


Generalization was then measured by leave-one-out cross-validation, with each colony predicted by a model trained on the remaining 80 colonies. Region- and boundary-based metrics remained concentrated in a high-accuracy regime (Figure~\ref{fig:f6}\hyperref[fig:f6]{b}), while dynamical measures separated propagation error, anisotropy deviation, and harmonic distortion more clearly (Figure~\ref{fig:f6}\hyperref[fig:f6]{c}). High mask accuracy was evaluated alongside the harder question of whether front motion and morphology remained dynamically consistent.

Trajectory-level analysis confirmed this distinction. Under a 50\% observation / 50\% prediction protocol, predicted effective radius and perimeter followed the measured time-aligned trajectories, and front propagation velocity reproduced the transition from rapid early expansion to later stabilization (Figure~\ref{fig:f6}\hyperref[fig:f6]{d--f}). The agreement persisted after prediction windows of different lengths were aligned by normalized horizon (Figure~\ref{fig:f6}\hyperref[fig:f6]{g}). Under the 80\% observation / 20\% prediction protocol, the same analysis resolved late-stage forecast evolution over a shorter window (Figure~\ref{fig:f6}\hyperref[fig:f6]{h}). Morpher therefore preserved not only occupied extent, but the temporal organization through which the front produced that extent.

\section{Discussion}

Bacterial swarming becomes biologically informative when it is read as a process of spatial commitment rather than as a final colony shape. Our results show that SM3 expansion is organized through the advancing interface, where local advances, arrests, protrusions, and branch rearrangements constrain which regions remain reachable and which sectors dominate later growth. For a gut-associated swarmer linked to mucosal repair, the relevant phenotype is therefore how this interface organizes future architecture before that architecture is fully expressed~\cite{de2021bacterial, pollackmilgate2024sm3, chen2021confinement, PhysRevE.109.064402}.

The weak separability of nominal assay conditions reinforces this view. Temperature, hydration, and agar-defined substrate mechanics set the arena in which swarming unfolds, but they did not specify reliable trajectories of future expansion. Colonies exposed to different conditions overlapped in morphological state space, while colonies in the same condition could diverge substantially (Figure~\ref{fig:f2}\hyperref[fig:f2]{a--d}). The environment still matters, but its influence is expressed through the interface that each colony builds under that environment. Local motility, interface geometry, and population structure can organize microbial spatial outcomes beyond what coarse environmental categories capture~\cite{gude2020bacterial, zhang2010collective, richter2024enhanced, xu2023geometrical}. The interface is therefore the state in which environmental constraint becomes colony-specific history.

This changes the meaning of segmentation. Morpher-S is not only a tool for producing cleaner masks. It defines the biological state on which temporal propagation operates. Once the colony edge is treated as the evolving variable, a boundary error becomes a state error. A smoothed protrusion, displaced rim, or contracted branch shifts the inferred location of future substrate encounter, microbial contact, and stress exposure. This effect was not subtle. A 0.67 percentage-point difference in mAP$_{50:95}$ between Morpher-S and YOLOv12, 92.48\% versus 91.81\%, expanded into a 2.4--3.1 IoU-point loss after autoregressive forecasting (Figure~\ref{fig:f3}\hyperref[fig:f3]{c} and Figure~\ref{fig:f4}\hyperref[fig:f4]{c}). Boundary fidelity is therefore part of biological state estimation, not image refinement. In this setting, segmentation determines which boundary history is allowed to become the predicted future~\cite{zhang2021deep, litjens2017survey, greener2022guide, li2025deep}.

The forecasting results separate interface prediction from ordinary future-frame synthesis. Generic video models could preserve coarse colony area, but they often smoothed protrusions, truncated lobe tips, or lagged behind the true interface (Figure~\ref{fig:f4}\hyperref[fig:f4]{a,b}). Those errors matter because the interface is where later spatial interactions are assigned. Morpher reached 95.42\% mIoU, 10.61 px HD$_{95}$, and 3.93 px ASSD under the 80\% observation / 20\% prediction protocol. Relative to SimVP+gSTA, this increased mIoU by 4.90 percentage points and reduced HD$_{95}$ and ASSD by 42.0\% and 55.7\%. The larger gains in boundary distances than in overlap show where the improvement lies. Morpher did not merely recover the correct occupied region. It preserved the moving interface through which local motion is organized into macroscopic expansion. Swarming prediction is therefore not image extrapolation alone, but preservation of the state that carries future spatial order~\cite{Chang2021MAU, wang2019mim, wang2017predrnn, wang2023predrnn, gao2022simvp, tan2023tau, tan2025simvpv2}.

The temporal ablations clarify why the interface is difficult to predict. Interface advance and interface structure were not the same objective. GRU achieved the lowest propagation RMSE, indicating strong short-term tracking of boundary displacement, but the Transformer with Morphon better preserved boundary fidelity and anisotropic organization (Figure~\ref{fig:f5}\hyperref[fig:f5]{b}). The hard part of swarming prediction is not radial motion alone. It is preserving the history by which branches, protrusions, and arrested sectors give the interface its present form. Morphon improved every backbone by retrieving earlier states, supporting a history-dependent view of interface evolution: similar instantaneous shapes can lead to different futures if they were produced by different sequences of advance, arrest, and deformation. This agrees with studies of expanding microbial frontiers, swarm development, and active interface patterning, where edge events can propagate into later population-scale structure~\cite{jeckel2023simultaneous, xu2023geometrical, hallatschek2007genetic, zdimal2025swarming}. The interface is therefore memory-bearing, not a static contour to be extrapolated.

The observation-ratio analysis shows when this memory becomes usable. At 50\% observation, prediction began near the transition from an early, morphologically similar phase to colony-specific divergence. Longer observation improved localization, as mIoU increased and ASSD decreased, but angular descriptors remained sensitive to the phase alignment of late branch rearrangements (Figure~\ref{fig:f5}\hyperref[fig:f5]{c}). Once the expansion mode became identifiable, additional history mainly refined the predicted position of the front. It did not remove uncertainty in the timing of anisotropic sector rearrangements. Leave-one-out evaluation showed that this behavior was not restricted to memorized trajectories. Morpher preserved the temporal organization of effective radius, perimeter, and front propagation velocity across held-out colonies (Figure~\ref{fig:f6}\hyperref[fig:f6]{b,d--h}). This is constrained predictability. Front geometry narrows the future without making living expansion deterministic.

This framework gives spatial microbiology a directly measurable motility state. Current computational and systems approaches increasingly recognize that microbiome function depends on spatial niches, host physiology, metabolic interactions, and local community organization~\cite{best2025metabolic, lotstedt2024spatial, ntekas2026spatial, prochazkova2024gut}. Swarming morphology adds the motile component of that spatial organization. It does not replace sequencing, metabolic modeling, microscopy, or gut-on-chip systems. It supplies an interface state through which movement, occupation, and environmental remodeling can be followed as a coupled process. This matters for bacteria whose swarming, mucus interaction, colonization, antimicrobial tolerance, and disease-associated behavior depend on where cells move and whom they encounter~\cite{pawul2024mucin, lai2009swarming, overhage2008swarming, piskovsky2023bacterial, sinha2025carbapenem}. A predictive interface state connects image-resolved motility phenotypes with spatial microbiome modeling.

The same logic can extend beyond SM3. We focused on SM3 because its swarming produces a well-defined surface-expanding phenotype that can be followed directly by time-lapse imaging, with a population-scale interface that branches, reorganizes, and carries predictive geometric information. These reported links give this phenotype a clear disease-relevant context~\cite{de2021bacterial, pollackmilgate2024sm3, chen2021confinement, PhysRevE.109.064402}. Other swarming species, mucus-rich substrates, mixed communities, and host-mimetic environments could be examined through the same interface-state lens. The projected colony front already provides a compact state variable, while measurements of colony height, cell density, nutrient gradients, chemical signaling, mucus structure, or host-associated microenvironments could enrich the state representation~\cite{pawul2024mucin, lee2025dynamics, prochazkova2024gut, riglar2019bacterial}. Mechanistic priors for active interfaces, surface wetting, nutrient access, and collective motility may further strengthen prediction under perturbation~\cite{beer2019statistical, bru2023swarming, jena2025spatio, xu2023geometrical}. Guided perturbation follows from the same state view. Predicted front trajectories could identify when and where substrate properties, chemical cues, neighboring populations, or engineered inputs are most likely to redirect swarming paths~\cite{doshi2023engineered, barbier2022engineering, finkelshtein2015bacterial}. In IBD-oriented assays, this would shift evaluation from final colony size to whether a perturbation steers repair-associated swarming toward favorable spatial reorganization.

Swarming becomes predictable when the advancing interface is treated as the evolving population state through which past spatial occupation constrains future expansion, rather than as a geometric outline of colony growth. The interface is where cells first meet new substrate, interact with neighboring populations, and carry forward the consequences of earlier advances, arrests, and branches. By resolving this interface and following its history, SwarmEvo and Morpher show that future colony architecture is already constrained before it is expressed at colony scale. For motile microbial populations, the advancing interface is therefore the population-scale state where local motion, environmental constraint, and boundary memory become future spatial organization.

\section{Methods}

\subsection{Swarming assay, imaging, and dataset construction}

The experimental workflow is shown in Figure~\ref{fig:f1}. A single SM3 colony was transferred from an LB agar plate into LB broth (10~g/L tryptone, 5~g/L yeast extract, and 5~g/L NaCl) and cultured overnight at \(37\,^{\circ}\mathrm{C}\) with shaking at 200~rpm. A \(5\text{--}8\,\mu\mathrm{L}\) aliquot was inoculated at the center of a freshly prepared swarming plate (LB with 0.5\%, 0.6\%, or 0.7\% agar; plate thickness 3--4~mm), incubated at \(30\,^{\circ}\mathrm{C}\) and \(\sim 90\%\) relative humidity for 4--6~h to activate swarming, and transferred to a time-lapse chamber maintained at \(27\,^{\circ}\mathrm{C}\), \(30\,^{\circ}\mathrm{C}\), or \(33\,^{\circ}\mathrm{C}\) and \(86\%\), \(90\%\), or \(94\%\) relative humidity. These temperature, humidity, and agar conditions span the permissive regime of SM3 swarming~\cite{pollackmilgate2024sm3, chen2021confinement, PhysRevE.109.064402}. Humidity was kept below the condensation threshold.

Images were acquired every minute with a vertically mounted high-resolution digital camera under uniform LED illumination until the colony reached the plate boundary or no further measurable expansion was observed. The Swarming Morphogenesis Evolution (SwarmEvo) dataset comprised 1,971 annotated images for segmentation and 276 long time series for temporal modeling. Recordings were stored at native resolution (\(1250\times1250\) px) with timestamps.

For downstream analysis, each sequence was converted into boundary-resolved colony masks using Morpher-S. These masks defined the forecasting state space. Training and validation partitions were split at the sequence level, with no colony contributing trajectories to both partitions. Dataset composition and augmentation procedures are provided in the Supplementary Information.

\subsection{Boundary-resolved interface-state construction}

Morpher-S was designed to recover colony fronts under two coupled imaging constraints: diffuse local boundaries and radial colony-scale organization. At the local scale, the swarming front often appears as a low-contrast boundary in which fine protrusions, branch tips, weak edge responses, and motion-related texture are intertwined. Texture--Edge Attention (TEA) preserves this boundary evidence through local depthwise filtering, multi-scale dilated texture encoding, and an edge-sensitive high-pass prior.

At the colony scale, expansion starts from an inoculation center and proceeds outward, while branches and rim fluctuations appear mainly as angular heterogeneity along the front. Polar--Context Attention (PCA) encodes this geometry by combining local features, large-kernel Cartesian context, and a polar branch operating in \((\rho,\theta)\) coordinates. TEA and PCA therefore recover weak local front evidence while preserving radial organization. Module formulations are provided in the Supplementary Information.

As shown in Figure~\ref{fig:f3}\hyperref[fig:f3]{a}, Morpher-S follows a prototype-based one-stage instance segmentation design~\cite{redmon2016yolo, bolya2019yolact}. A five-stage hierarchical convolutional backbone interleaves TEA and PCA blocks during progressive downsampling. Multi-scale features are fused by a PANet-style bidirectional neck~\cite{lin2017feature, liu2018path}. Prediction is performed at feature levels \(P3\), \(P4\), and \(P5\) through heads for class scores, bounding boxes, and mask coefficients. A lightweight Protonet produces \(k=32\) shared prototypes, which are linearly combined with the predicted coefficients, cropped, and thresholded to generate final colony masks.

Training used the segmentation objective
\[
\mathcal{L}_{\mathrm{seg}}
=
\lambda_{\mathrm{box}} \mathcal{L}_{\mathrm{box}}
+
\lambda_{\mathrm{cls}} \mathcal{L}_{\mathrm{cls}}
+
\lambda_{\mathrm{dfl}} \mathcal{L}_{\mathrm{dfl}}
+
\lambda_{\mathrm{mask}} \mathcal{L}_{\mathrm{mask}}.
\]
Here \(\mathcal{L}_{\mathrm{box}}\), \(\mathcal{L}_{\mathrm{cls}}\), \(\mathcal{L}_{\mathrm{dfl}}\), and \(\mathcal{L}_{\mathrm{mask}}\) denote box, classification, distribution focal, and mask losses. Loss weights were fixed across experiments.

\subsection{Forecasting interface-state evolution}

Morpher-F forecasts swarming expansion in mask space rather than image space. Each input mask sequence is encoded by a shared multi-scale spatial encoder into framewise latent descriptors \(z_t\in\mathbb{R}^{256}\). Intermediate encoder features are retained for decoding, and sinusoidal temporal encodings are added before temporal modeling.

As shown in Figure~\ref{fig:f5}\hyperref[fig:f5]{a}, Morpher-F contains a multi-scale spatial encoder, a temporal sequence model, a Morphon memory block, and a multi-scale decoder. Observed masks are compressed into a latent sequence and future states are predicted in latent space. In autoregressive inference, each decoded prediction is re-encoded and used as the next input. Morphon retrieves observed front history by cross-attention with a learnable query derived from the aggregated observation state and injects the retrieved structural memory through a learnable gate \(\alpha\in(0,1)\). The decoder reconstructs predicted masks while reinjecting encoder features to preserve protrusions and fine curvature.

Forecasting was autoregressive unless otherwise stated. After observing \(T_{\mathrm{obs}}\) frames, the model predicted the next latent state, decoded it into a mask, re-encoded the prediction, and repeated this process for subsequent steps. Parallel prediction was used only in matched ablations.

The temporal module was instantiated as a vanilla RNN~\cite{elman1990rnn}, GRU~\cite{cho2014gru}, LSTM~\cite{hochreiter1997long}, or Transformer encoder~\cite{vaswani2017attention}. All variants used the same encoder--decoder backbone, latent dimensionality, and past-to-state formulation. Recurrent variants used three recurrent layers, and the Transformer used three encoder blocks.

Prediction was supervised at each forecast step by
\[
\mathcal{L}_{\mathrm{pred}}
=
\frac{1}{T}\sum_{t=1}^{T}
\left(
\lambda_{\mathrm{foc}} \mathcal{L}_{\mathrm{foc}}^{(t)}
+
\lambda_{\mathrm{iou}} \mathcal{L}_{\mathrm{iou}}^{(t)}
+
\lambda_{\mathrm{bd}} \mathcal{L}_{\mathrm{bd}}^{(t)}
\right).
\]
Here \(\mathcal{L}_{\mathrm{foc}}\), \(\mathcal{L}_{\mathrm{iou}}\), and \(\mathcal{L}_{\mathrm{bd}}\) denote focal, soft IoU, and boundary losses. Loss weights were fixed across experiments. Implementation details, sequence construction, and baseline-matched comparisons are provided in the Supplementary Information.

\subsection{Training and implementation}

For Morpher-S, images were resized to \(640\times640\). Training used the Ultralytics YOLO framework with SGD, an initial learning rate of \(6\times10^{-3}\), three-epoch linear warm-up, decay to 1\% of the initial rate, momentum \(=0.937\), weight decay \(=5\times10^{-4}\), batch size \(=16\), mixed precision, and early stopping after validation saturation.

For Morpher-F, binary masks from a single segmentation model were uniformly subsampled with a fixed stride and split into observation and prediction segments at ratios of \(0.5/0.5\), \(0.6/0.4\), \(0.7/0.3\), \(0.8/0.2\), and \(0.9/0.1\). Masks were resized to \(640\times640\). Training used AdamW with an initial learning rate of \(5\times10^{-5}\), weight decay \(=10^{-4}\), batch size \(=2\), 300 epochs, 10\% warm-up followed by cosine annealing, gradient clipping at 1.0, and mixed precision. Model selection used the highest validation mIoU. Fixed random seeds and deterministic backend settings were used, with TensorFloat-32 acceleration enabled where available.

All baseline models were retrained under matched preprocessing and identical train/validation splits. Model-specific settings for YOLOv11/12, SAM/SAM2, MAU, MIM, PredRNN, PredRNNv2, SimVP+TAU, and SimVPv2+gSTA are provided in the Supplementary Information.

\subsection{Evaluation}

Segmentation was evaluated by mAP$_{50:95}$, image-wise IoU, and Dice coefficient. Forecasting was evaluated by spatial, dynamical, and angular metrics. Spatial fidelity was measured by mIoU, HD, HD$_{95}$, and ASSD. Dynamical fidelity was measured by radial-velocity RMSE and the Temporal Consistency Index (TCI). Angular organization was measured by \(|\Delta\mathrm{NAS}|\) and \(|\Delta \mathrm{H}_{2}|\).

In the main text, mAP$_{50:95}$ is the primary segmentation metric, while HD$_{95}$ and ASSD are the primary forecasting metrics because region overlap can remain high even when the contour drifts. Formal metric definitions are provided in the Supplementary Information.

\section*{Acknowledgements}
This work was supported by the National Natural Science Foundation of China under Grant No. 12202275 and by the Shanghai Jiao Tong University Explore X Fund.

\section*{Competing interests}
The authors declare no competing interests.

\section*{Data availability}
All data supporting the findings of this study are publicly available. 
The Swarming Morphogenesis Evolution (SwarmEvo) dataset, including both the segmentation and temporal prediction subsets, is hosted at 
\url{https://huggingface.co/datasets/ShengyouDuan/SwarmEvo}. 
The dataset provides polygon-based segmentation annotations and time-lapse sequences for bacterial swarming experiments, and is released to support reproducibility and further research in morphology-aware modeling.

\section*{Code availability}
The complete implementation of the proposed framework, including Morpher-S for morphology-aware segmentation and Morpher-F for autoregressive temporal forecasting, is publicly available at \url{https://github.com/ShengyouDuan/From_shape_to_fate__making_bacterial_swarming_expansion_predictable}. The repository contains all training and evaluation code required to reproduce the results reported in this work.

\bibliographystyle{naturemag}
\bibliography{reference}

\clearpage

\clearpage

\clearpage

\clearpage

\clearpage

\clearpage

\end{document}